\documentclass[12pt]{iopart}
\usepackage{graphicx,ulem}
\usepackage{booktabs}
\usepackage{caption}
\usepackage{dsfont,hyperref}
\usepackage{textcomp}
\usepackage[framemethod=tikz,xcolor=true,nobreak=true]{mdframed}
\usepackage{multicol}

\usepackage{abstract}
\renewcommand{\abstractname}{}    
\usepackage{xcolor}

\usepackage{comment}
\usepackage{iopams}  
\newcommand\ddfrac[2]{{\displaystyle\frac{\displaystyle #1}{\displaystyle #2}}}

\begin{document}

\title{Determining intrinsic sensitivity and the role of multiple scattering in speckle metrology}
\author{Morgan Facchin$^1$, Saba N. Khan$^1$, Kishan Dholakia$^{1,2,3}$ and Graham~D.~Bruce$^1$} 
\address{$^1$SUPA School of Physics and Astronomy, University of St Andrews, North Haugh, St Andrews KY16 9SS, UK}
\address{$^2$School of Biological Sciences, University of Adelaide, Adelaide, South Australia, Australia}
\address{$^3$Centre of Light for Life, University of Adelaide, Australia}
\ead{gdb2@st-andrews.ac.uk}
\vspace{10pt}
\begin{indented}
\item[]\today
\end{indented}
\markboth{}{}

\begin{abstract}

Speckle patterns are a powerful tool for high-precision metrology, as they allow remarkable performance in relatively simple setups. Nonetheless, researchers in this field follow rather distinct paths due to underappreciated general principles underlying speckle phenomena. Here, we advise on a universal metric of intrinsic speckle sensitivity, and on the advantages and disadvantages of multiple scattering. This will catalyse progress in speckle metrology but will also translate to other domains of disordered optics which are undergoing rapid developments at present.
\end{abstract}

\hfill

Speckle patterns are the intensity patterns resulting from the random interference of light. Despite being commonly associated with loss of information, they have given rise to a wide range of applications, such as imaging at depth inside tissue \cite{bertolotti2022imaging,yoon2020deep} and around corners \cite{faccio2020non}; widely-used clinical techniques for assessing blood flow at depth \cite{Briers13}; astronomical observations such as imaging of stars through the atmosphere \cite{labeyrie1970attainment} or the first measurements of the supermassive blackhole at the centre of our galaxy \cite{ghez1998high}; controlling the properties of a new generation of lasers \cite{sapienza2022controlling,cao2019complex} and photonic devices \cite{cao2022harnessing}; and optical computation and information processing \cite{gigan2022imaging}. 

\begin{figure}[hb!] 
\centering\includegraphics[width=\textwidth]{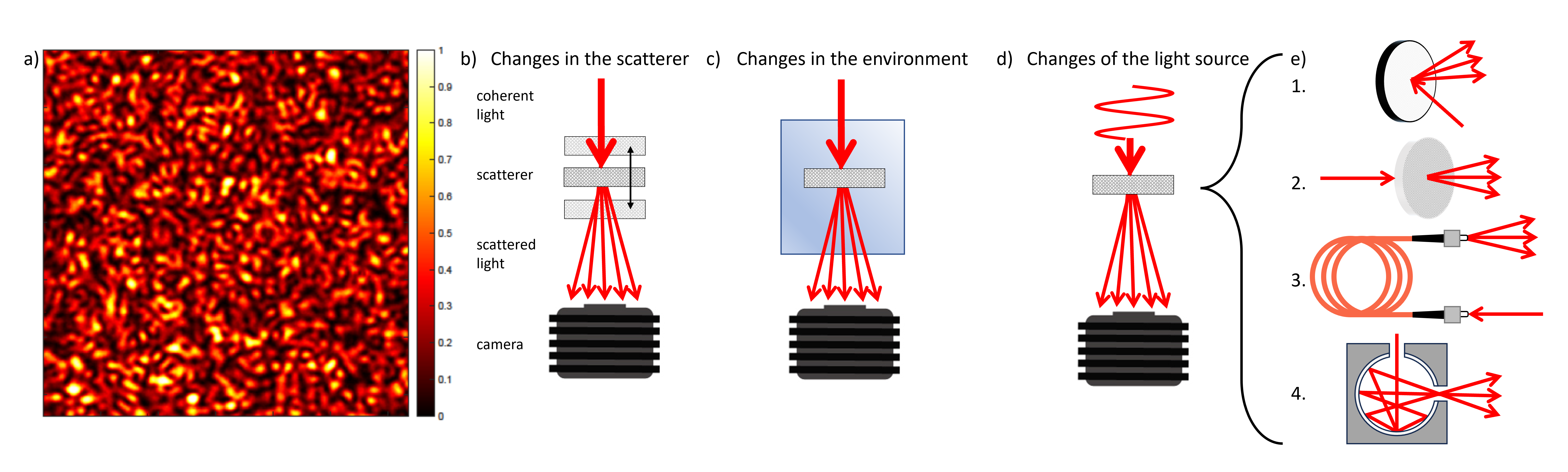}
\captionsetup{font=footnotesize,width=0.8\textwidth}
\caption{\textbf{Overview of speckle metrology.} \textbf{a} A typical speckle pattern, resulting from the randomisation of coherent light. The specific speckle pattern generated is influenced by variations in \textbf{b} the random medium, \textbf{c} the surrounding environment, or \textbf{d} the illuminating light. \textbf{e} Examples of optical elements that have been used to produce the speckle pattern: 1. reflective scatterer, 2. ground glass, 3. multi-mode fibre, 4. integrating sphere.}
\label{fig:overview}
\end{figure}

Another important field of application that is seeing increasing activity is metrology. Changes in the speckle pattern can be used to probe changes in the scattering medium, in the environment in which the medium is embedded, or in the illuminating light (FIG.~\ref{fig:overview}). For example, the displacement or rotation of a test scattering surface can be measured by looking at the global translation of the speckle pattern reflected by the surface \cite{archbold70,wang2006core,wang2006vortex,wang2005phase}. The same scheme also allows the measurement of wavelength variations of a test laser beam impinging on a rough surface at an angle, where a change in wavelength yields a global translation of the resulting speckle \cite{chakrabarti2015speckle,hanson2018dynamic}. Similar measurements can be made using speckles produced by a plane transmissive diffuser \cite{Mazilu12,sun2023near}, or by optical time domain reflectometry \cite{wan2020high}. The method can be extended to spectrometry \cite{redding2012using,redding2013compact,redding2013all,Redding14b,coluccelli2016optical,cao2017perspective,wan2021review,inalegwu2023machine}, and hyperspectral imaging \cite{kurum2019deep}, with varying setup schemes and analysis techniques. Refractive index can be determined by measuring the translation \cite{Guo18,luo2023refractive} or morphological change \cite{Trivedi19} of a speckle pattern under the influence of a test material. The polarisation state of a laser beam can be retrieved from the speckle pattern it produces, using a set of reference speckles for known polarisation states \cite{facchin2020pol}, which can be extended to spatial polarimetry \cite{kohlgraf2009spatially}. Other applications include measurements of strain and deformation \cite{murray2019speckle,graciani20223d,warren2022multimode}, vibration \cite{bianchi14}, speech and heartbeat \cite{zalevsky09}, or the dynamics of drying paint \cite{van2016paint}. \textcolor{black}{The fact that speckle can be used to measure changes in so many measurands is both a blessing and a curse, and a significant new direction is ensuring the stability of the speckle patterns against unwanted effects by engineering the scattering medium \cite{Sun22,cai2023compact} or adapting the analysis approach used \cite{sun2023near,cai2023compact,bruce19}. Applications are now being targeted outside the academic research lab, such as smart-beds in care settings \cite{warren2022multimode} and industrial spectrum analysers \cite{kelley2023high,murray2023high}, and this has seen the development of increasingly portable techniques including building systems on smartphone platforms \cite{tan2023optical}. }

Implementations of speckle metrology are being pursued with a wide variety of scattering media, including transmissive diffusers \cite{Mazilu14}, reflective rough surfaces \cite{chakrabarti2015speckle}, photonic chips \cite{redding2013compact}, multimode \cite{redding2012using} and single-mode fibres \cite{wan2020high}, integrating spheres \cite{Facchinmodel}, biological tissue \cite{Briers13}, and even everyday materials such as sellotape \cite{malone2023diffuserspec}. Researchers also use different analytical approaches to extract meaning from the speckle, including transmission matrices \cite{cao2017perspective}, supervised \cite{Metzger17} and unsupervised machine learning \cite{gupta19}, compressive sensing \cite{kurekci2023single} and temporal analyses \cite{graciani20223d,xu2023cavity}. 

One striking recent trend in speckle metrology is the emergence of techniques that use speckle patterns resulting from the multiple scattering of light. Examples of such geometries include nano-engineered photonic crystals \cite{redding2013compact,Sun22}, propagation through a succession of plane diffusers \cite{Tran20}, multiple reflections inside an integrating sphere \cite{Metzger17,davila2020single}, or propagation in a multimode fibre \cite{redding2012using,Redding14,Wan15,Liew16,Bruce20}. The speckle patterns produced in those ways have proved to be much more sensitive to the measurand of interest (i.e. changing more rapidly with a change in the measurand). This is most strikingly shown in the case of: (i) Wavelength, where attometre-scale fluctuations can now be measured \cite{bruce19,Facchinmodel,gupta19} (6 orders of magnitude improvement over a single scattering surface \cite{chakrabarti2015speckle}); (ii) Displacement, where picometer-scale resolutions have been obtained (3 orders of magnitude improvement over previous speckle-based methods \cite{wang2006core}); and (iii) Refractive index, where $10^{-9}$ resolution has been achieved \cite{Facchin2021ref} (3 orders of magnitude improvement over previous speckle-based methods \cite{Tran20}). Multiple scattering seems to come as a universal solution to improve performance in speckle metrology. However, is this the case for all possible measurands? 

The answer is no. Here, we outline a general framework to understand why. We highlight a metric to quantify the intrinsic sensitivity of a speckle pattern to a measurand of interest, which is independent of the analysis ultimately applied to the speckle patterns. We then use this metric to provide a criterion to decide whether multiple scattering is indeed beneficial for a given metrology task, and illustrate this with experimental examples. 

\section*{A metric for sensitivity  }

The performance of a given speckle metrology technique ultimately depends on two factors: the intrinsic sensitivity of the speckle pattern to the measurand of interest, and the analysis method applied. Here, we will not compare the various analysis methods, but rather discuss the intrinsic sensitivity of the speckle pattern. To quantify such sensitivity, we first need a metric to quantify the difference between two given speckle images.  

Two metrics are typically found in the literature, denoted as the Spectral Correlation Function $SCF(\Delta\alpha)$ \cite{redding2012using,redding2013all,Redding14,bruce19} and Similarity $S(\Delta\alpha)$ \cite{Facchinmodel,Facchin2021ref,facchin_displace,Tran20,hanson2018dynamic,chakrabarti2015speckle,Trivedi19,trivedi2022opto}, for a given change $\Delta\alpha$ in the measurand of interest $\alpha$. Although the $SCF$ is typically used in the context of wavelength change, as its name suggests, we use it here for any generic measurand. These are defined as

\begin{equation} \label{correl}
SCF(\Delta\alpha)= \bigg\langle  \frac{ \big\langle I I'\big\rangle_{\alpha}}
{\big\langle I\big\rangle_{\alpha} 
\big\langle I'\big\rangle_{\alpha}} 
-1 \bigg\rangle_x
\end{equation}

\begin{equation} \label{simil}
S(\Delta\alpha)= \bigg\langle  \bigg( \frac{I - \langle I\rangle_x}{\sigma_I} \bigg)\bigg( \frac{I'-\langle I'\rangle_x}{\sigma_{I'}} \bigg) \bigg\rangle_x,
\end{equation}

\noindent with $I$ and  $I'$ two speckle images (before and after a change in $\alpha$), using the shorthands $I=I(x,\alpha)$ and  $I'=I(x,\alpha+\Delta\alpha)$, where $x$ is position on the speckle image, $\langle\cdot\rangle_{x}$ is averaging over the speckle image, $\langle\cdot\rangle_{\alpha}$ averaging over $\alpha$, and $\sigma$ is the standard deviation of the speckle image (i.e. $\sigma_I^2=\langle I^2\rangle_{x} - \langle I\rangle_{x}^2$).

These two metrics of speckle change can now be used to infer speckle sensitivity. Indeed both metrics, when plotted against $\Delta\alpha$, show a globally decreasing curve. The Half Width at Half Maximum (HWHM) of this curve provides a natural measure of sensitivity of the speckle pattern with respect to $\alpha$. This is the sense in which \textit{sensitivity} shall be used in the rest of this article. 

\begin{figure}[h!] 
\centering\includegraphics[width=0.7\textwidth]{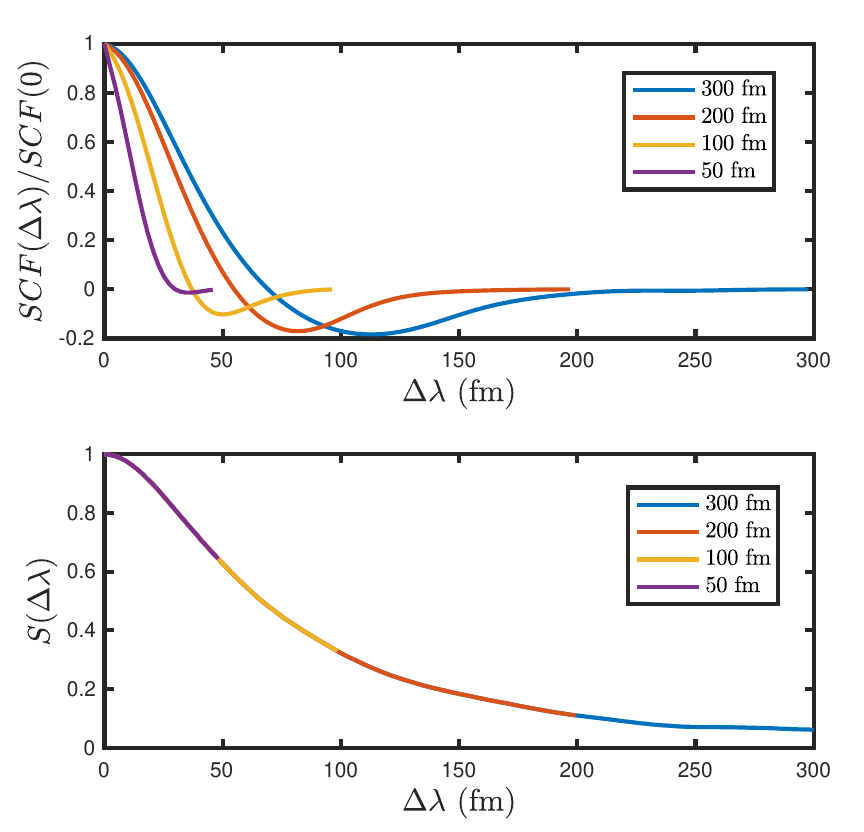}
\captionsetup{font=footnotesize,width=.8\textwidth}
\caption{\textbf{A comparison of metrics.}  \textbf{a} Spectral Correlation Function $SCF(\Delta\lambda)$ and \textbf{b} Similarity $S(\Delta\lambda)$, computed using different subsets of the same dataset, for a speckle pattern varying as a result of a wavelength change $\Delta\lambda$. We use subsets with a total range of 300 fm, 200 fm, 100 fm, and 50 fm. We find that the shape of $SCF(\Delta\lambda)$ changes in each case, while $S(\Delta\lambda)$ does not.    }
\label{crapcoef}
\end{figure}

We shall demonstrate however that $SCF(\Delta\alpha)$ should be used with particular care, as its HWHM depends strongly on the range of $\alpha$ in the data from which it is calculated. To illustrate this, we use experimental data obtained with 780 nm laser light scattered by an integrating sphere (see details in Supplementary Information). The resulting speckle is set to change by applying a wavelength variation of 300 fm. Then, we compute both $SCF(\Delta\lambda)$ and $S(\Delta\lambda)$ using the full 300 fm range of data, or using only a subset of it, i.e. only the first 200 fm, 100 fm, and 50 fm. This is equivalent to a shorter applied wavelength variation in experiment. The resulting profiles of $SCF(\Delta\lambda)$ and $S(\Delta\lambda)$ are shown in FIG. \ref{crapcoef}. 

We find that the width of $SCF(\Delta\lambda)$ changes significantly when the range of $\Delta\lambda$ changes. This dependency is due to the presence of averaging over $\alpha$ ($\lambda$ in this case) in EQ. \ref{correl}. On the other hand, $S(\Delta\lambda)$ is essentially independent of the range of $\Delta\lambda$. Similar behaviour is also seen with different experimental data and even simulated speckles. Therefore, if $SCF(\Delta\alpha)$ is employed, its computation should encompass a substantial range of variation, approximately ten times larger than the apparent HWHM of the curve. This variability causes difficulties for the reliable comparison of different experimental platforms. Researchers can consider using $S(\Delta \alpha)$ as a more robust metric, and if $SCF(\Delta \alpha)$ is used then the range of values of $\alpha$ used in the calculation must be provided. Note that in FIG. \ref{crapcoef}, $SCF(\Delta\alpha)$ is normalised to $SCF(0)$, as the amplitude also varies depending on the subset used. 

An optimal metric of speckle sensitivity should be independent of the range of values on which speckle change is investigated. Exploring larger variations of $\alpha$ should only extend the graph of our function of interest, and not change the portion already obtained. Beyond this practical point, $S(\Delta\alpha)$ also has the advantage of being directly interpretable: it is the Pearson correlation coefficient computed between two speckle images, which naturally quantifies the similarity (used here in the common sense of the term) of the two images. Also, the simple expression of the Pearson correlation coefficient lends itself to a direct analytical estimation based on first principles, as we shall see in the next sections.

Before moving on, we clarify an important point regarding this metric. The width of the similarity curve acts as a metric to assess the sensitivity of a scattering medium to the measurand of interest. It does not, in general, directly indicate the precision limit. When a single parameter is being measured, the precision is determined by the smallest detectable speckle variation, generally dictated by camera noise. Only when multiple examples of the same measurand are being measured, does the width of the similarity curve (often referred to as the ``speckle correlation limit") indicate a limit to the resolution with which we can distinguish them. For example, in the context of wavelength measurement - the speckle correlation limit indicates the resolution with which we can distinguish two peaks in a spectrum, but does not indicate the precision with which we can measure a single wavelength of a monochromatic source \cite{bruce19}. The relationship is similar to the diffraction limit in imaging - this restricts our ability to distinguish between neighbouring emitters but not our ability to determine the position of an isolated emitter, a fact that underpins key super-resolution microscopy methods \cite{palm,storm}.

\section*{A general model of speckle sensitivity}
Before proceeding to the key point of this article on the utility of multiple scattering in speckle metrology, we review a general approach used to model speckle change for an arbitrary transformation. This leads us to our criterion for whether multiple scattering is advantageous for a given metrological application.
Using the Similarity (EQ. \ref{simil}) as a measure of speckle change, and simplifying our notations with brackets now meaning only averaging over the image, we have

\begin{equation} \label{eq:simil_I}
S= \bigg\langle  \bigg( \frac{I_{} - \langle I_{}\rangle}{\sigma_I} \bigg)\bigg( \frac{I_{}'-\langle I_{}'\rangle}{\sigma_{I'}} \bigg) \bigg\rangle,
\end{equation}
with, as before, $I$ and $I'$ the speckle images before and after a change in our measurand of interest. In the following the prime superscript shall indicate the quantity after the change. 

\textcolor{black}{Now a fully developed speckle is assumed}, i.e. one that is the superposition of a large number of independent waves, each having a uniform phase distribution on a $2\pi$ interval (see \cite[chap~2]{Goodman} for a more precise definition). This is a good model when light is sufficiently randomised, which is the case when light is reflected on a sufficiently rough surface, and almost certainly the case when multiple scattering is involved. Assuming this, $S$ can be expressed as \textcolor{black}{\cite[chap~3.3.4]{Goodman}}

\begin{equation} \label{simil_E}
    S=\left |  
    \frac{\Big\langle \boldsymbol{E}^\dagger \boldsymbol{E}'\Big\rangle}
    {\sqrt{\Big\langle  \boldsymbol{E}^\dagger \boldsymbol{E}  \Big\rangle} \sqrt{\Big\langle  \boldsymbol{E}'^\dagger \boldsymbol{E}'  \Big\rangle}  } 
    \right |^2, 
\end{equation} 
where $\boldsymbol{E}$ and $\boldsymbol{E'}$ are respectively the underlying fields of the speckle before and after the change, where $\dagger$ denotes the conjugate transpose. The underlying field is the electromagnetic field $\boldsymbol{E}$ such that the observed intensity follows $I\propto |\boldsymbol{E}|^2$. The field $\boldsymbol{E}$ is a 3$\times$1 complex-valued vector representing either the electric or magnetic component of the wave. The quantity inside the absolute square of EQ. \ref{simil_E} can be seen as the (complex) Pearson correlation of the field itself. 

The output field $\boldsymbol{E}$ can now be expressed as a linear transformation of the incident field $\boldsymbol{E}_0$, reading $\boldsymbol{E}=T\boldsymbol{E}_0$, where $T$ is a linear operator. We can always write this, even if $T$ cannot be determined in practice (see for example \cite{matthes2019optical}), as long as no non-linear processes are involved. This notation is useful to consider speckle change in the most general way. In the following, we examine the two possible cases where the change occurs in the input field, and in the scattering medium.   

\subsection*{A change in the input field} \label{beam}
In the case of a change in the input field (e.g. its amplitude/phase/polarisation profile), we have $T'=T$, $\boldsymbol{E}=T\boldsymbol{E}_0$, and $\boldsymbol{E}'=T\boldsymbol{E}'_0$. With this, EQ. \ref{simil_E} becomes

\begin{equation} \label{scattering}
    S=\left |  
    \ddfrac{\Big\langle \boldsymbol{E}_0^\dagger T^\dagger T \boldsymbol{E}'_0 \Big\rangle}
    {\sqrt{\Big\langle  \boldsymbol{E}^\dagger_0  T^\dagger T\boldsymbol{E}_0  \Big\rangle} \sqrt{\Big\langle  \boldsymbol{E}'^\dagger_0  T^\dagger T\boldsymbol{E}'_0  \Big\rangle}  } 
    \right |^2.
\end{equation}

\noindent The quantity in the absolute square is close to the correlation of the input field $\boldsymbol{E}_0$, except for the presence of the $T^\dagger T$ operator (when applied to an operator, $\dagger$ represents the adjoint). However we can show that this operator is approximately proportional to the identity operator ($\mathds{1}$) under quite general assumptions. We can show this in two extreme cases: 

\begin{enumerate}
    \item First, consider an ideal geometry where the speckle is produced by a single reflection on a rough surface and collected in the Fraunhofer region. We can write $T\propto\mathcal{F}e^{i\phi}$, where $\phi$ is a phase mask equivalent to the rough surface and $\mathcal{F}$ is the Fourier transform operator. The $T^\dagger T$ term becomes $(\mathcal{F}e^{i\phi})^\dagger (\mathcal{F}e^{i\phi})= e^{-i\phi} \mathcal{F}^\dagger \mathcal{F} e^{i\phi} = e^{-i\phi} \mathds{1} e^{i\phi}= \mathds{1}$, where we use the fact that the Fourier transform is a unitary operator ($ \mathcal{F}^\dagger \mathcal{F}=\mathds{1}$). 
    \item Second, consider a more complex geometry involving multiple scattering, where light is highly randomised. In this case it is more convenient to adopt a discrete description of the system, where $\boldsymbol{E}_0$ and $\boldsymbol{E}$ are expressed not as 2-dimensional functions of space, but as N-dimensional vectors containing the values of the field at N given points of space (in a scalar model for simplicity) \textcolor{black}{\cite{matthes2019optical}}. This corresponds to a practical situation, where the field is measured at $N$ pixels. In this description $T$ is now a matrix (commonly known as the transmission matrix \cite{popoff2010measur_transm}), and each element of $T^\dagger T$ can be seen as the covariance between two columns of $T$. If the geometry is complex enough, we expect $T$ to be a complex random matrix with uncorrelated columns, such that $T^\dagger T$ is approximately proportional to $\mathds{1}$ (now representing the identity matrix). 
\end{enumerate}
In any case lying somewhere between those two extremes, the argument is less clear and care must be taken. Assuming $T^\dagger T\propto \mathds{1}$, EQ. \ref{simil_E} becomes
\begin{equation} 
    S=\left |  
    \ddfrac{\Big\langle \boldsymbol{E}^\dagger_0  \boldsymbol{E}'_0 \Big\rangle}
    {\sqrt{\Big\langle  \boldsymbol{E}^\dagger_0  \boldsymbol{E}_0  \Big\rangle} \sqrt{\Big\langle  \boldsymbol{E}^\dagger_0   \boldsymbol{E}'_0 \Big\rangle}  }\right |^2, \label{similsame}
\end{equation}
which is the absolute square of the incident field's correlation. In other words, the Similarity becomes independent on the properties of the scattering medium. It follows that the width of the Similarity curve (i.e. the sensitivity of the speckle) cannot be changed by a careful choice of diffusing geometry, such as multiple scattering. 

\subsection*{A change in the scattering medium}

In the case of a change in the scattering medium (e.g. its geometry or optical properties), we have $\boldsymbol{E}=T\boldsymbol{E}_0$ and $\boldsymbol{E}'=T'\boldsymbol{E}_0$, and EQ. \ref{simil_E} becomes

\begin{equation} 
    S=\left |  
    \ddfrac{\Big\langle \boldsymbol{E}^\dagger_0 T^\dagger T' \boldsymbol{E}_0 \Big\rangle}
    {\sqrt{\Big\langle  \boldsymbol{E}^\dagger_0  T^\dagger T\boldsymbol{E}_0  \Big\rangle} \sqrt{\Big\langle  \boldsymbol{E}^\dagger_0  T'^\dagger T'\boldsymbol{E}_0  \Big\rangle}  } 
    \right |^2.
\end{equation}


This can be made more explicit by expressing the field on the observation plane as the superposition of the fields coming from all possible paths through the scattering medium \textcolor{black}{\cite{Facchinmodel,facchin2023speckle}}. In a discrete description, a path can be defined as a succession of discrete elements of the scattering medium between which light propagates in straight lines. In this description, it can be shown (see Supplementary Information \textcolor{black}{for a simplified derivation}) that the Similarity becomes 

\begin{equation} \label{Tp}
S =  \left |  \ddfrac{   \sum_{p}   \sqrt{t_{p}t_{p}'} e^{i \Delta\phi_p}  }
{\sqrt{\sum_{p}  t_{p}\sum_{p}  t_{p}'}       } \right |^2,
\end{equation}
where $t_p$ is the power transmission of path $p$ (such that if a power $P$ goes through path $p$, a power $t_p P$ reaches the observation plane), and $\Delta \phi_p = \phi_p'-\phi_p$ the phase variation on path $p$. The phase on a given path is given by $\phi_p=nkz_p$, with $n$ the refractive index, $k$ the wavenumber, and $z_p$ the length of path $p$. 

Here we see that the situation is reversed: the Similarity no longer depends on the incident field $\boldsymbol{E}_0$, but only on properties of the scattering medium. The dependency on the scattering medium is reflected by the presence of the $t_p$ and $t_p'$ (transmission properties) and  $\Delta\phi_p$ (geometry of the system, as it includes $z_p$). 

\subsection*{A general criterion: path-dependent vs path-independent effects}
 
In deriving equations (\ref{similsame}) and (\ref{Tp}), we find that the properties of the scattering medium only come into play when considering a path-dependent effect. That is, when the effect changes the amplitude or phase of light in a way that depends on its path through the scattering medium.

This allows us to express our criterion for determining whether multiple scattering increases speckle sensitivity to a given effect: multiple scattering can increase speckle sensitivity only for a path-dependent effect. For a path-independent effect, multiple scattering makes no difference.

Path-dependent effects generally correspond to changes in the scattering medium, and path-independent effects generally correspond to changes in the incident light. There is only one exception, which is a change in wavelength. This is because of the presence of wavelength (through the wavenumber $k$) in the exponential term of EQ. \ref{Tp}. This exception is why path-dependency is a better criterion than our initial approach of light vs scattering medium, although this works in most cases. 

An exhaustive list of path-dependent effects can be obtained by direct inspection of EQ. \ref{Tp}. These are a change in transmission, refractive index, wavelength, or deformation of the medium, by the presence of $t$, $n$, $k$, and $z$, respectively. These have been studied in the literature with multiple scattering, except for transmission, for which only an analytical study is given in the context of gas absorption \cite[Chapter~3.6.1]{facchin2023speckle}. Note however that EQ. \ref{Tp} does not include path-dependent polarisation effects (such as the Faraday effect or birefringence). Such effects have not been studied in the literature. 

Note that this criterion applies to what we may call ``natural" speckle patterns, i.e. speckle patterns that are only determined by the medium used and not by any form of control from the experimenter. A range of techniques involving wavefront shaping exist where the experimenter can tailor the speckle patterns to exhibit particular properties \cite{carpenter2015observation,matthes2021learning,ambichl2017super,han2023tailoring,bender2023spectral,bouchet2021maximum,bender2022coherent}, in which case some assumptions of the model (such as fully developed statistic) may not be met. 

\section*{Experimental examples}
To illustrate our criterion, we use experimental data where the speckle pattern is produced using the following scattering media: 

\begin{itemize}
\item a diffusely reflecting disk, as in \cite{archbold70,wang2006core,wang2006vortex,wang2005phase,chakrabarti2015speckle,hanson2018dynamic,facchin2020pol}
\item a multimode fibre, as in \cite{redding2012using,inalegwu2023machine,murray2019speckle,warren2022multimode,Redding14,Wan15,Liew16,Bruce20}
\item an integrating sphere, as in  \cite{graciani20223d,facchin_displace,Metzger17,davila2020single,Facchinmodel,gupta19,Facchin2021ref,Boreman90}
\end{itemize}

\noindent For each of these scattering media, we apply two particular transformations: a change in polarisation (path-independent), and a change in wavelength (path-dependent). 

A change in polarisation is a path-independent effect, which according to our criterion should yield a speckle change that is independent of the scattering medium. For an input beam whose polarisation is linear and rotated by an angle $\theta$, EQ. \ref{similsame} gives the Similarity (see Supplementary Information) 
\begin{equation}
S=\cos^2\theta, \label{eq:cosangle}    
\end{equation}

\noindent which most strikingly exemplifies independence on the scattering medium.

\begin{figure}[h!] 
\centering\includegraphics[width=1.0\textwidth]{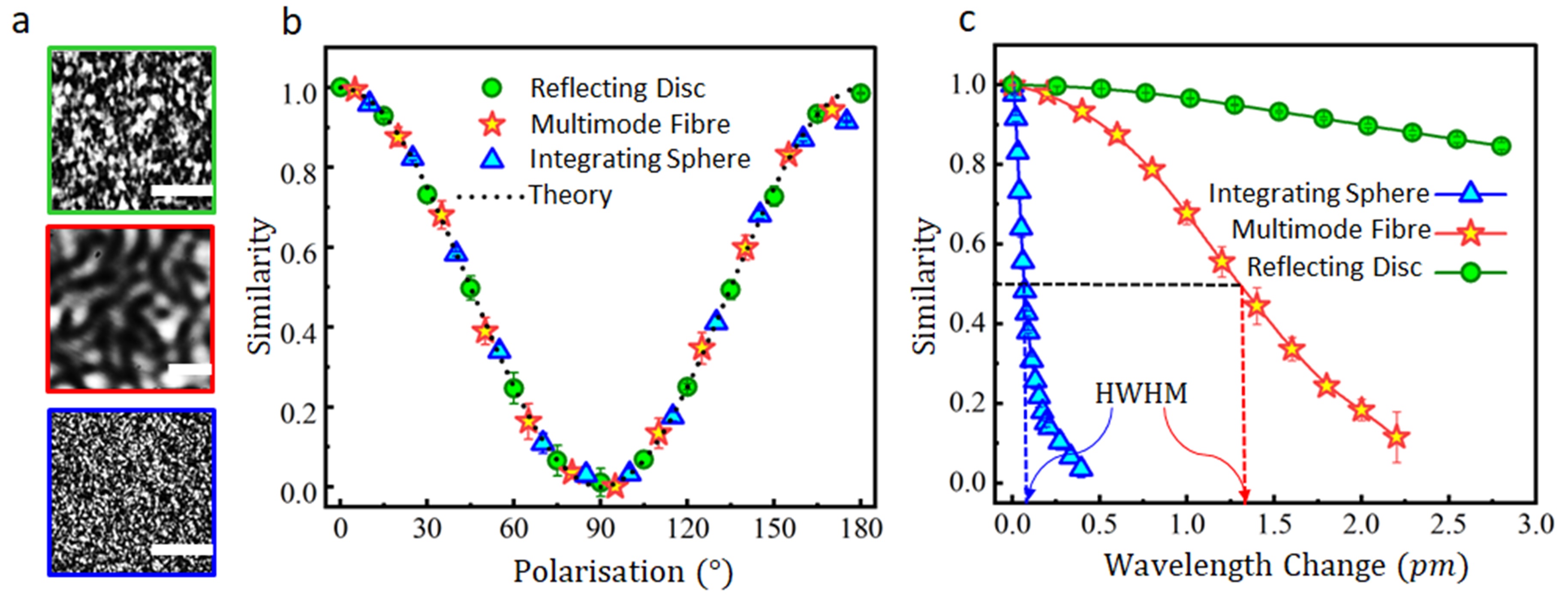}
\captionsetup{font=footnotesize,width=.8\textwidth}
\caption{\textbf{Speckle sensitivity to different changes.} \textbf{a} Examples of speckle patterns from a reflecting disk (top, green), a \textcolor{black}{step-index} multimode fibre (middle, red) and an integrating sphere (bottom, blue) showing very different morphologies. Scale bar is 1 mm in each case. \textbf{b} Similarity as a function of polarisation angle for each scattering medium, and predicted profile (dotted black). As polarisation change is a path-independent effect, the three profiles are almost identical. \textbf{c} Similarity as a function of wavelength change for each scattering medium. As wavelength is a path-dependent change, the 3 profiles are remarkably different. }
\label{fig:exp}
\end{figure}

The obtained Similarity profiles are shown in FIG. \ref{fig:exp}, where we see that the three Similarity profiles are identical within experimental error and agree with EQ. \ref{eq:cosangle}, despite the speckle patterns having markedly different morphologies in each case. 

A change in wavelength, unlike a change in polarisation, produces a path-dependent phase shift and therefore the choice of scattering medium should affect the resulting Similarity profile. Previous analytical studies confirm this: 

\begin{itemize}
    \item In the case of a single rough surface, the explicit solution for the Similarity is shown in~\cite{chakrabarti2015speckle}, and is relatively complex. Its HWHM follows $\Delta \lambda_0 \approx  \lambda^2 /(2\pi w)$, with $\Delta \lambda_0$ the HWHM, and $w$ the $1/e$ width of the incident beam, assuming both incidence and observation angle equal to 45°. 

   \item In the case of a multimode fibre, the explicit solution for the Similarity is not known, however it was shown \cite{redding2013all,Rawson80} that its HWHM follows $\Delta \lambda_0 \propto \, \lambda^2/(L\, \textrm{N}\!\textrm{A}^2)$, with $L$ the fibre's length, and $\textrm{N}\!\textrm{A}$ its numerical aperture.

   \item In the case of an integrating sphere,it was shown in \cite{Facchinmodel} that the Similarity is a Lorentzian function whose HWHM is give by $\Delta \lambda_0=3\lambda^2\left|\ln\rho\right|/(8\pi R)$, with $\rho$ the reflectivity of the sphere's inner surface, and $R$ its radius.
\end{itemize}

The obtained Similarity profiles are shown in FIG.~\ref{fig:exp}. The Similarity profiles have indeed very different widths: the HWHM is 6.3 pm for the rough surface, 1.3 pm for the multimode fibre, and 66 fm for the integrating sphere. The integrating sphere has a sensitivity 20 times larger than the multimode fibre and 95 times larger than the single rough surface. 

\subsection*{Why is multiple scattering efficient for path-dependent effects?}

In this section we discuss in more details the reason why multiple-scattering increases sensitivity for path-dependent effects. Where intuition might suggest that the increased total path-length is the origin of the enhanced sensitivity, it is in fact the increased \textit{spread} in path-length distribution \textcolor{black}{\cite{Facchin2021ref}}. 

For any given path-dependent effect, the transformation of light on a given path increases with the length of that path. For example, a wavelength change induces a phase change given by $  \Delta k \, z$ (assuming $n=1$), with $\Delta k$ the corresponding wavenumber change, and $z$ the length of the path considered. This effect is indeed directly proportional to path-length. However, to infer the resulting speckle change, we must consider the interference of the different paths. If all paths acquire the same phase, however large it may be, the resulting speckle will not change at all. Consider then two particular paths that interfere on the observation plane, the phase difference between them is $\Delta k \, \Delta z$, with $\Delta z$ their length difference. The presence of $\Delta z$ shows that geometries with higher path-length differences will produce more sensitive speckles. \textcolor{black}{Therefore, as multiple scattering geometries typically imply higher path-length differences, their speckle sensitivity tends to be higher.} \textcolor{black}{This is analogous to interferometers with unequal path lengths, where sensitivity is increased~\cite{narasimha2007fully}.} 

\begin{figure}[h!]
\centering\includegraphics[width=0.7\textwidth]{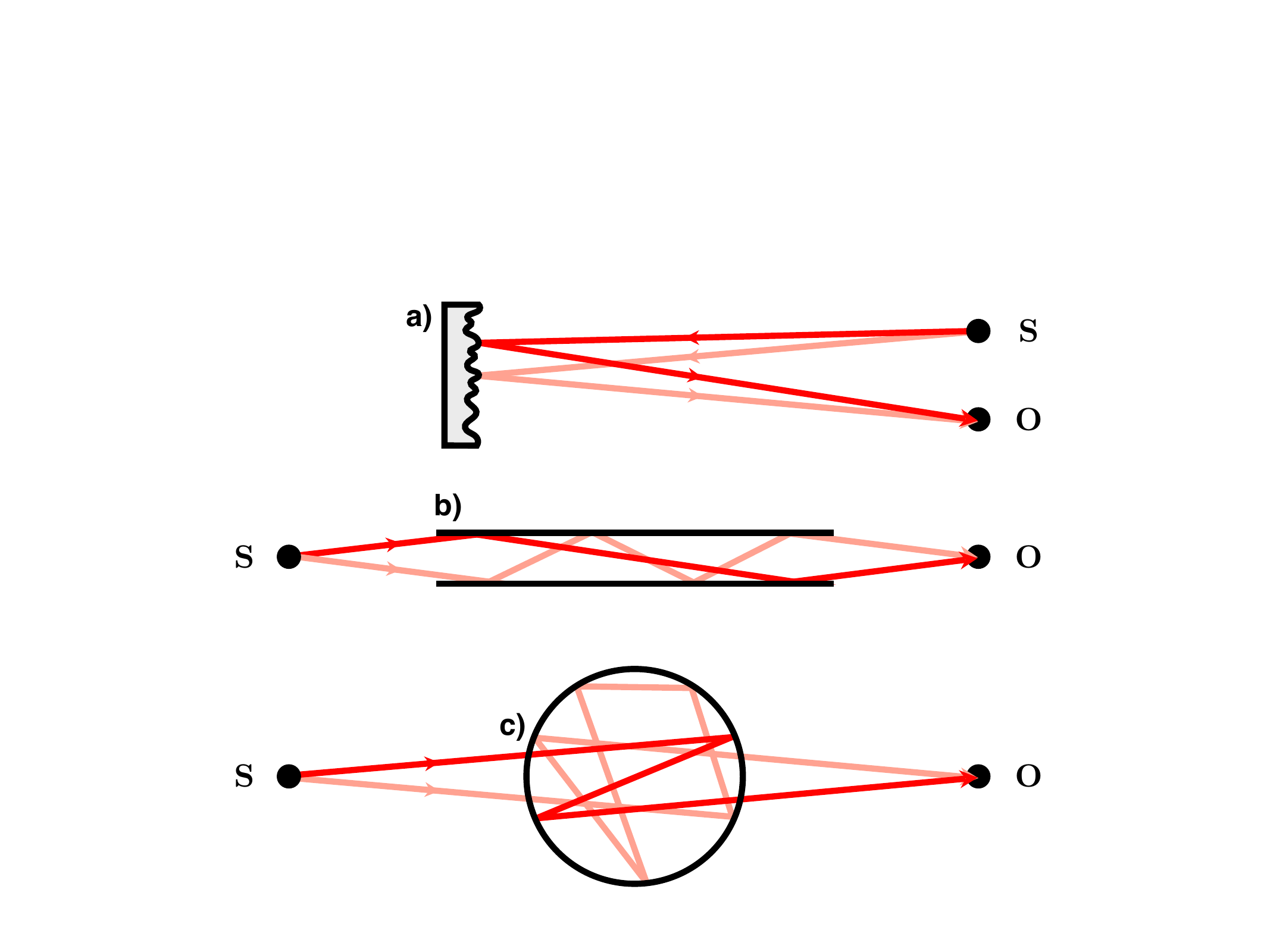}
\captionsetup{font=footnotesize,width=.8\textwidth}
\caption{\textbf{Path length distributions in typical scattering elements.} \textbf{a}  a reflecting surface. \textbf{b} a multimode fibre. \textbf{c} an integrating sphere. In each case two possible paths are represented, between the source (S) and a point of the observation plane (O). The spread in path-length distribution increases from top to bottom. }
\label{media}
\end{figure}

This is schematically illustrated in FIG. \ref{media} for the three scattering media considered in our experimental examples, where different possible paths are shown. For a single reflective surface, the amount of path-length difference is determined by the area accessible to light, which is consistent with the $w$ dependence of the HWHM. For a multimode fibre (adopting a ray-optics point of view where propagation is seen as the multiple reflections of light in a cylindrical reflector) we can see geometrically that light entering at higher angles yields a wider range of lengths, proportionally to the length of the fibre itself. This is consistent with the NA and $L$ dependence of the HWHM, respectively. Finally, for an integrating sphere the length of each path is proportional to the sphere's radius, and consequently also the spread in path-length, which is consistent with the $R$ dependence of the HWHM. 
 
\section*{Summary and Conclusion}

Speckle can be implemented for a variety of metrological purposes. But how should one design an optimal speckle sensor for a particular purpose? In this context, our initial suggestion is a standardisation of metrics that will allow for comparison between scattering media presented in the literature, namely to use the Similarity (or Pearson correlation coefficient). 

Using this metric, we showed that the most relevant criterion regarding speckle sensitivity is whether the effect under study is path-dependent (i.e. changes the amplitude/phase/polarisation of light differently depending on the path through the scattering medium) or path-independent (i.e. changes the the amplitude/phase/polarisation of light in the same way for all paths). For a path-independent effect, speckle sensitivity depends only on the properties of the input beam. On the other hand, for a path-dependent effect, the situation is reversed: speckle sensitivity depends only on the properties of the scattering medium. Therefore, multiple scattering only increases sensitivity for path-dependent effects. For path-independent effects, multiple scattering makes no difference.

Therefore, it is essential that the scattering medium for a given speckle application must be chosen according to this criterion. Multiple scattering is not always advantageous, and can in fact be detrimental. Indeed multiple scattering geometries can involve higher absorption, leading to thermal expansion, whose effect on the speckle pattern can dominate the effect under study \cite{Facchin2021ref,facchin_displace}. On the other hand, much higher optical throughput can be obtained with single scattering surfaces, as compared to an integrating sphere for example. Also, vibrations generally couple more weakly to a single scattering geometry than they do to a multiple scattering geometry. Therefore, for path-independent effects, or path-dependent effects where high sensitivity is not desired, single scattering geometries may be the optimal experimental design. On the other hand, for path-dependent effects where high sensitivity is required, multiple scattering geometries are optimal. 

\section*{Data availability} 
The data underpinning this work will be available through the University of St Andrews Research Data Portal at \href{https://doi.org/10.17630/bcb2bff1-0d20-454a-89c5-8fd64b9e8ff8}{https://doi.org/10.17630/bcb2bff1-0d20-454a-89c5-8fd64b9e8ff8}. 
\ack
This work was supported by funding from the Leverhulme Trust (RPG‐2017‐197), the UK Engineering and Physical Sciences Research Council (EP/P030017/1,EP/R004854/1), and the Australian Research Council (FL210100099). We acknowledge useful discussions with Hui Cao on the subtleties of the comparison between Spectral Correlation Function and Similarity metrics, as well as Nicolas Dubost, Paul Hawthorne, and Trisha Bhide.

\section*{References}

\bibliography{sample}

\end{document}